# AUTHENTICATION DEVICES IN FOG-MOBILE EDGE COMPUTING ENVIRONMENTS THROUGH A WIRELESS GRID RESOURCE SHARING PROTOCOL

Tyson Brooks

Syracuse University, USA

## ABSTRACT

*The rapid growth of the Internet of Things (IoT), cloud computing, Fog computing, mobile edge computing and wireless grids has resulted in the widespread deployment of relatively immature technology. These technologies, which will primarily use 5G wireless communication networks, are becoming popular because they can be deployed quickly with little infrastructure and lends themselves to environments utilizing numerous internet connected devices (ICD). There are, however, many significant challenges faced by security designers, engineers and implementers of these networks in ensuring that the level of security afforded is appropriate. Because of the threat of exploitation, these networks have to be protected by a robust security architecture due to these technologies being plagued with security problems. The authentication of smart ICDs to IoT networks is a critical mechanism for achieving security on these new information system platforms. This article identifies an authentication process required for these ICDs, which will need to prove their identity to authenticate to an IoT fog-mobile edge computing (FMEC) cloud network through a wireless grid authentication process. The purpose of this article is to hypothesize a generic authentication methodology for these FMEC clouds uses in an IoT architecture. The proposed methodology, called wg-IoT, must include the integration of Fog computing, wireless grids and mobile edge computing clouds to create this new IoT architecture. An authentication process developed from the resource sharing protocol (RSP) from a wireless grid is first developed and proposed for the authentication of ICDs. The wireless grid core components must be embedded in IoT devices or sensors depending on their capability to handle five primary functions: management of identification [ID] and presence, permissions management, data transferability, application-programming interface [API] and security.*

## 1. INTRODUCTION

The IoT paradigm, which consist of a network of embedded sensor connected to the Internet, is rapidly gaining ground in wireless telecommunications (Zhang et al. 2012). However, only recently has the IoT market begun to experience rapid growth. This is due to several factors, which include: the extensive spread of the Internet, next generation 5G networks, high penetrations of mobile devices usage, new cloud computing platforms, fog-mobile edge computing (FMEC) and microbrowsers[1,2]. Industry analysts predict that the global IoT market will reach $772.5B in 2018, attaining a Compound Annual Growth Rate (CAGR) of 14% through the 2017-2021 forecast period - surpassing the $1 trillion mark in 2020 and reaching $1.1 trillion

---

[1] https://www.transparencymarketresearch.com/mobile-edge-computing-market.html
[2] https://www.ibm.com/developerworks/library/wa-browse/index.html

in 2021[3]. The Fog computing market is expected to reach $617.3M by 2025[4] and over $5.6B devices will utilize mobile edge computing, primarily in the manufacturing, energy, and transportation industries[5]. All of these market forces are converging, which in turn is making the IoT market increasingly data centric.

Any cloud network is subject to becoming the target of exploitation by individuals or groups outside of the authorized group of intended users and/or devices (Firdous et al. 2017). In addition, the advances in wireless technology have given the common individual the ability to establish a communications connection with anyone at any time. This ability has created a general addiction to connectedness and has opened new opportunities for the exploitation of communications. The exploitation of a cloud network by adversaries is conducted for four main reasons; (1) to gain access to data flowing over the network, (2) to disrupt the flow of data on the network, (3) to parasitically usurp the networks resources (e.g. to use a network free of charge), and (4) provide disinformation by injecting false data into the network in order to mislead and to cause confusion and doubt (Gupta & Gupta, 2018; Ahmed et al. 2018). However, this situation has now changed, as ICDs will be able to perform these exploitations with limited human interactions.

With these assumptions as background, authentication of [wireless] ICDs in these FMEC clouds must involve strong encryption to prevent eavesdropping and must involve mutual authentication to ensure that sensitive information is transmitted only over legitimate FMEC cloud networks. Authentication will require that ICDs prove their identity through a hardware/software token, a challenge/response mechanism or a combination of these and other methods of identification. Authentication should be two-way, the ICD must authenticate to the network, and the network should authenticate to the ICD. This lets the network know that it is communicating with a valid, non-malicious ICD. By having the network authenticate to the ICD, it is less likely that an unauthorized ICD will be able to pose as valid in the network. The purpose of this article is to hypothesize a generic authentication methodology for ICDs in FMEC clouds utilizing the IoT architecture.

## 2. FOG COMPUTING

Cisco, Inc. defines the Fog as 'an extension of the cloud to be closer to the things that produce and act on IoT data, processing IoT data closer to where it is produced and needed solves the challenges of exploding data volume, variety, and velocity and accelerating awareness and response to IoT events by eliminating a round trip to the cloud for analysis' (Cisco 2015b). The National Institute of Standards and Technology (2018) recently defined Fog computing as:

A layered model for enabling ubiquitous access to a shared continuum of scalable computing resources. The model facilitates the deployment of distributed, latency-aware applications and services, and consists of fog nodes (physical or virtual), residing between smart end-devices and centralized (cloud) services. The fog nodes are context aware and support a common data management and communication system. They can be organized in clusters - either vertically (to support isolation), horizontally (to support federation), or relative to fog nodes' latency-distance to the smart end-devices. Fog computing minimizes the request-response time from/to supported applications, and provides, for the end-devices, local computing resources and, when needed, network connectivity to centralized services.

---

[3] https://www.networkworld.com/article/3244927/internet-of-things/new-idc-report-forecasts-huge-growth-for-iot.html
[4] https://www.grandviewresearch.com/press-release/global-fog-computing-market
[5] https://www.thorntech.com/2017/11/edge-computing-and-the-cloud-future-of-iot/

The characteristics of the Fog (proximity and location awareness, geo-distribution, hierarchical organization) make it the suitable platform to support both energy-constrained wireless grids (Bonomi et al. 2012). However, this implies a number of characteristics that make the Fog a non-trivial extension of the cloud including edge location, location awareness, and low latency (Zhu et al. 2013). The Fog, for instance, will play an active role in delivering high quality streaming to moving vehicles, through proxies and access points positioned along highways and tracks. Localization of data processing is a fundamental and essential issue for operational wireless sensor networks (WSNs) (Sheu et al. 2008).

A large number of Fog nodes, as a consequence of the wide geo-distribution, as evidenced in sensor networks in general and the Smart Grid in particular will provide support for mobility (Mukherjee et al. 2017; Ekanayake et al. 2018). It is essential for many Fog applications to communicate directly with mobile devices, and therefore support mobility techniques, such as the Locator ID Separation Protocol (LISP) 1, that decouple host identity from location identity, and require a distributed directory system (Zhu et al. 2013). Support for on-line analytic and interplay with the Cloud, the Fog is positioned to play a significant role in the ingestion and processing of the data close to the source (Zhu et al. 2013). Figure 1 presents the idealized information and computing architecture supporting the future IoT applications, and illustrates the role of Fog computing through compute, storage, and networking resources building blocks of both the Cloud, the Fog and the mobile edge of the network.

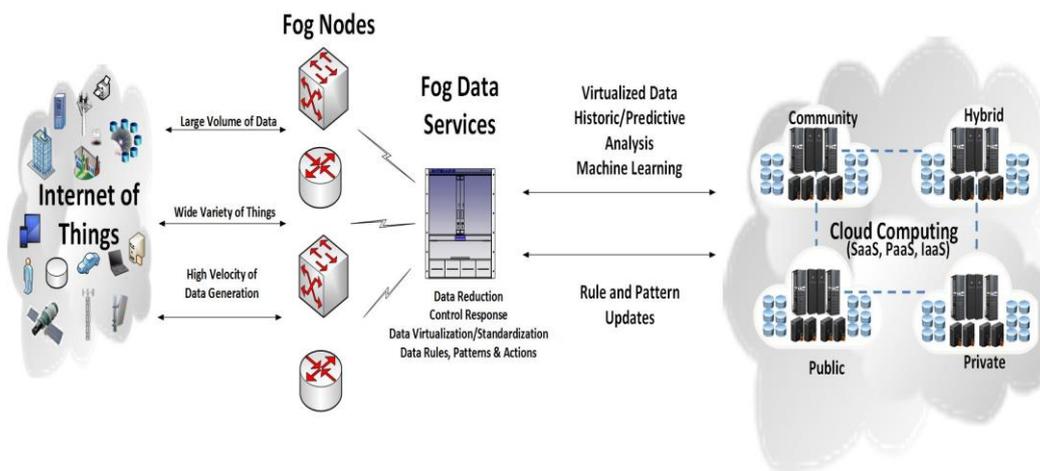

Figure 1. The Internet of Things and Fog Computing
(Source: Cisco 2015a; Brooks & McKnight, 2017)

## 3. WIRELESS GRIDS

As displayed in Figure 2, a wireless grid is an augmentation of a wired grid that facilitates the exchange of information and the interaction between heterogeneous wireless devices (Agarwal et al. 2004). The IoT is an integrated part of the future Internet including existing and evolving Internet and network developments and could be conceptually defined as a dynamic global network infrastructure with self-configuring capabilities (Vermesan et al. 2011). This is based on standard and interoperable communication protocols where physical and virtual smart "things/objects" (e.g. wireless sensors, actuators, radio frequency identification [RFID] tags, smart mobile devices, mobile robots, etc.) have identities, physical attributes, and virtual personalities, use intelligent interfaces, and are seamlessly integrated into the information network (Vermesan et al. 2011). The application of wireless grids, FMEC and IoT architectures is a key objective which facilitates information sharing and provides a means for the system to get information (whether wired or wirelessly) to individuals to satisfy their needs.

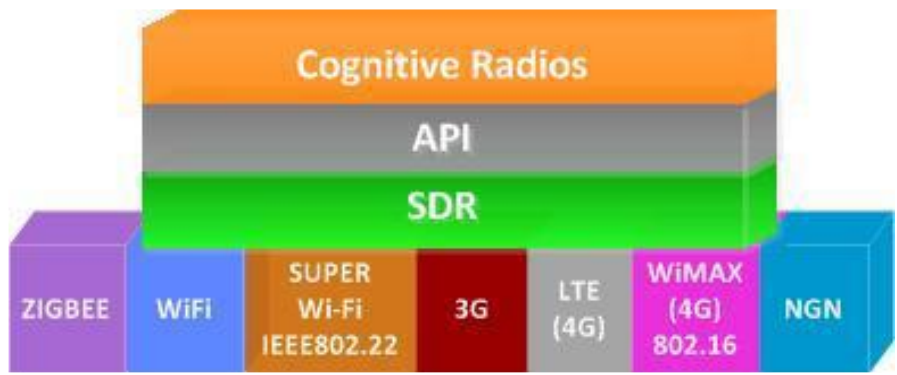

Figure 2. WIGIT Open Framework
(Source: McKnight et al. 2015)

In the IoT, smart "things/objects" (i.e., internet connected devices or ICDs) are expected to become active participants in business, information and social processes where they are enabled to interact and communicate among themselves and with the environment by exchanging data and information sensed about the environment (Vermesan et al. 2011; Uckelmann et al. 2011). Services will be able to interact with these smart ICD's using standard interfaces that will provide the necessary link via the Internet, to query and change their state and retrieve any information associated with them, considering security and privacy issues (Vermesan et al. 2011). 'Things' can only become context aware, sense, communicate, interact, exchange data, information and knowledge if they are suitably equipped with appropriate object-connected technologies; unless of course they are human 'things' or other entities with these intrinsic capabilities (Brooks, 2017). In this vision, using intelligent decision-making algorithms in software applications, appropriate rapid responses can be given to physical phenomena, based on the very latest information collected about physical entities and consideration of patterns in the historical data, either for the same entity or for similar entities (Yan et al. 2010; Vermesan et al. 2011).

However, the application of authentication of these ICD's within this environment brings about new and challenging problems from an information security perspective. Traditional security mechanisms, such as identification/authentication and access control (authorization) are complicated in these environments, requiring new standards and the development of new products. Counter-intuitively, the more wireless grids, FMECs and IoT architectures that exist, the more vulnerable they will towards cyber-attacks. This is because, with wireless grids, FMECs and IoT architectures, every user/device/thing/object may have the right to access the system causing errors and flaws. For this reason, individual user/device/thing/objects will continue to be the primary targets of malicious software (or malware) attacks.

## 4. WIRELESS GRID 'EDGEWARE'

As displayed in Figure 3, Edgeware, a new class of software applications, enables the ad hoc connection of people, devices, software and services in a personal cloud, supported by personal cyber infrastructure (Brooks et al. 2013, McKnight et al. 2015). Edge devices are routers, switches, routing switches, integrated access devices (IAD), multiplexers, and a variety of metropolitan area network (MAN)/wide area network (WAN) access devices that provide entry points into enterprise or carrier/service provider core networks which translate between one type of network protocol and another (McKnight et al. 2015). Edgeware applications can dynamically make use of content and resources present in devices - phones, laptops, PCs, cameras, printers, screens, etc. – through connectivity via a wireless grid (McKnight et al. 2015).

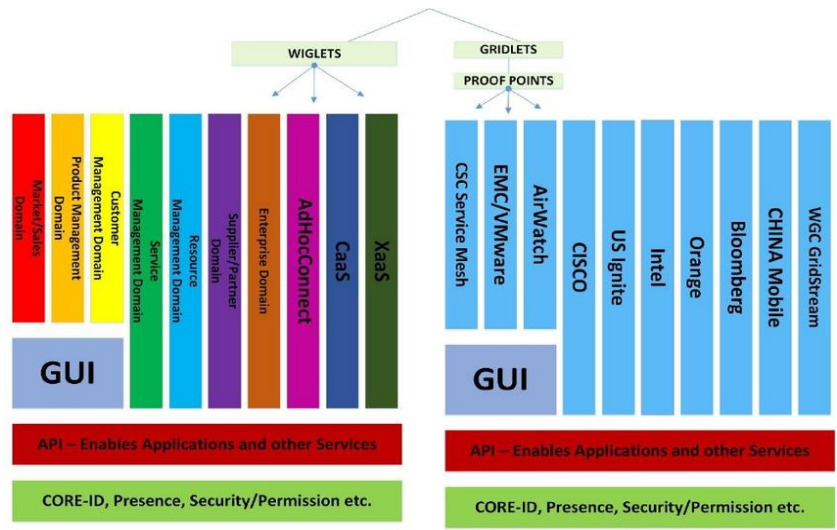

Figure 3. Edgeware Grid Core
(Source: McKnight, ed. WiGiT v0.3)

The blue boxes on the right in Figure 3 represent Edgeware applications that sit on a user interface, which in turn sits on an API [McKnight et al. 2015]. These may represent dozens or hundreds of different sorts of mini- programs that enable different kinds of resource sharing and functionality. Edgeware applications are typically delivered as a service; and come in 2 primary varieties: Gridlets, that is, proprietary Edgeware applications, and Wiglets, that is, non-proprietary open Edgeware applications (McKnight et al. 2015). Not all devices enabled on a wireless grid need to have an Edgeware application sitting on them to be accessible and active.

The only thing that must be deployed for a wireless grid to work is for the Grid Core to be on some intelligent machine, somewhere, with rights to control other 'edge' resources such as sensors that may not have the capability to have the core components installed; which may be facilitated by one or more Gridlets and/or Wiglets (McKnight et al. 2015). Other network hardware, software, services, and content may be controlled and shared through the wireless grid 'Edgeware' as these may not be or cannot become self- aware devices on the grid; However, if those 'edge' resources are in a relationship with other hardware, software, and services, which are part of the wireless grid, they may function as if they were fully cognitive (McKnight et al. 2015) A further differentiation in the varieties of Edgeware applications may also be drawn between peer-to-peer implementations, and cloud to edge applications, which may appear at first glance to be a basic client-server implementation. In both cases, however, the Edgeware application may be able to interact dynamically with other types of Edgeware applications. Meaning, the architecture and open specifications presented here allow for ad hoc, peer-to-peer applications and services to interact with cloud services (McKnight et al. 2015).

The Grid Core components are represented by the green box and embedded in certain devices or sensors depending on their capability, which makes every device a node on the wireless grid (McKnight et al. 2015). This core is extremely 'light' and easy to embed on a wide range of different kinds of equipment. McKnight et al. (2015) identified that users are allowed to share and manage the digital resources at their fingertips through applications of the architecture's eight core components:

- the authentication and authorization component [AAC]
- the billing, accounting and charging component [BAC] which provides access to the things/objects financial information,
- the messaging and presence component [MPC] which provides scalable messaging, manages the availability of a thing/object and the method or language of communication with that thing/object,
- the metadata component [MC] which creates, edits, and generally manages the metadata for an ICD, (5) the resource management component [RMC] which is responsible for aggregating and searching metadata about things/objects within the context of authentication and works closely with the AAC and MPC,
- the economic and legal policy component [ELP] which supports economic and legal policies,
- the communication protocols component [CPC] which is a sub-system that manages the interaction with specific types of resources, such as printers, files, etc. and is needed to interact within a wireless grid, identifies and manages network and internetwork communications including IP and other protocols (e.g. Bluetooth), provides connections with other wireless grids and across the internet, and,
- the security component.

The wireless grid architecture core components handle four primary functions, which make the grid-enabled ecosystem possible: management of identification (ID) and presence, permissions management, data transfer ability, and API/interfacing (McKnight et al. 2015). The layers above the core are comprised of the API which enables connections with other applications and services, the User Interface (which may or may not be necessary depending on the device upon which it sits), and finally the Edgeware applications are shown in blue in Figure 2.

Once a grid is established then resources can be published or accessed across the grid, enabling the infinite functional possibilities of the Grid technology. There are three classes of wireless grid applications (McKnight et al. 2015):

- Class 1: Applications aggregating information from the range of input/output interfaces found in nomadic and mobile devices,
- Class 2: Applications utilizing the locational and contextual characteristics in which the devices will be found and,
- Class 3: Applications leveraging the mesh network capabilities of groups of devices: Workplace-As-A-Service (WPaaS), Compute-Infrastructure-As-A-Service (CIaaS) and Virtual Private Cloud (VPC) are reference architectures that potentially can meet the unique requirements to satisfy enterprise-grade customers.

These wireless grid Edgeware systems give the user greater mobility and flexibility. Although these characteristics of services provide a number of advantages especially towards its integration into IoT architectures, they leave these wireless grids Edgeware devices vulnerable to attacks and other security related problems (Brooks et al. 2013). It is exactly the mobile nature of the devices that exposes it to greater risk of data loss or theft and this contrasts with the wired service, which terminates in one location, such as in the home or office, making it more safeguarded (Brooks et al. 2013). The trend of integrating these new complex systems with advanced computer and communication technologies has introduced serious cyber-security concerns, especially in these new architectures; where the environment will no longer be regarded as reliable to support communications as before (Brooks et al. 2013). For example, due to the important role of the smart grid as the key energy IoT infrastructure, the information infrastructure needed to route data by providing the dynamic ad-hoc sharing of heterogonous devices and the need to protect its information security is an extremely important task, which can significantly contribute to security issues given the threat of cyber-attacks (Li et al. 2012). Traditional security mechanisms, such as identification/authentication and access control (authorization) are complicated in these new environments, requiring new standards and the development of new security products.

Due to its physical broadcast nature, these new communication networks are generally more vulnerable to malicious and accidental threats than their wired counterparts. As a result of this inherent vulnerability, security is a mandatory component. While it's more difficult and potentially more important to secure this communication, the issues, threats and the respective required services to adequately respond to these threats are mostly the same for wired and wireless technology. Alternatively, the task of providing security services for these networks is more complicated than in wired networks. Power and bandwidth limitations, often non-existent in wired networks, impose considerable constraints on the complexity and efficiency of security protocols.

## 5. FOG MOBILE EDGE COMPUTING (FMEC)

A mobile grid combines mobile computing and grid computing and develops rapidly (Zeng et al. 2008). FMEC can be mobile, portable or fixed. However, in general, the mobile user unit is a mobile, wireless device. These user devices provide one basic function – connectivity with an access point or a base station providing mobile services. Samimi et al. (2006) define FMECs as clouds that support autonomic communication at the wireless edge of the Internet, defined as those nodes that are one, at most a few, wireless hops away from the wired infrastructure. FMEC enable dynamic instantiation, composition, configuration, and reconfiguration of services on an overlay network to support mobile computing (Samimi et al. 2006). FMEC provides a distributed infrastructure designed to facilitate rapid prototyping and deployment of services that enhance communication performance, robustness, and security and include a collection of low level facilities that can be either invoked directly by applications or used to compose more complex services (McKinley et al. 2006).

As displayed in Figure 3, the FMECs model supports dynamic composition and reconfiguration of services to support clients at the wireless edge and provide an infrastructure for composing autonomic communication services (Samimi et al. 2006; Vermesan et al. 2011). FMECs (see Figure 3) allows information systems to mirror the integrated, evolving business processes of an enterprise to deliver specific capabilities and service levels. The ability of FMECs to change, evolve, and manage business processes throughout an enterprise is changing the way information technology development, integration and deployment works. Pervasive FMECs in an enterprise will identify and highlight cross-functional dependencies and encourage cooperation and communication between and among functional units and information technology.

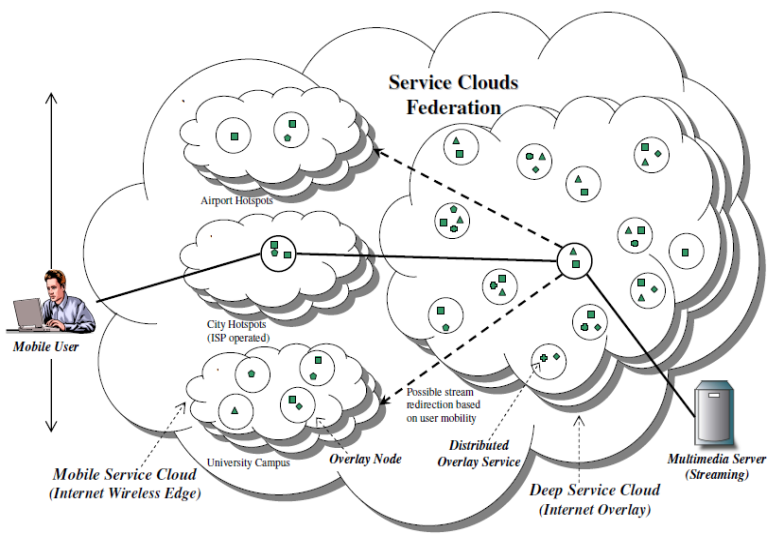

Figure 4. Fog Mobile Edge Computing
(Source: Samimi et al. 2006)

Wireless grids are infrastructure-less mobile ad-hoc networks that can intelligently and dynamically interconnect users and stakeholders at multiple sites, transfer digital media, assume and respond to different equipment types, and adapt to low power conditions and diminished communications capabilities (McKnight et al. 2004). There are two modes of wireless grid creation, user mode and machine-based mode, as displayed in Figure 5 and Figure 6, compare a 'human-user'-centric grid with a 'node-based' grid (McKnight et al. 2015). In purely conceptual terms, it is evident that in both cases the outermost frontier of what is currently possible (i.e., engaging the full range of user types) with device heterogeneity considered on an infinite axis only goes so far; the promise of the wireless grid technology is the capability of 'M2M' communication via a virtual distributed operating system that enables the IoT (McKnight et al. 2015).

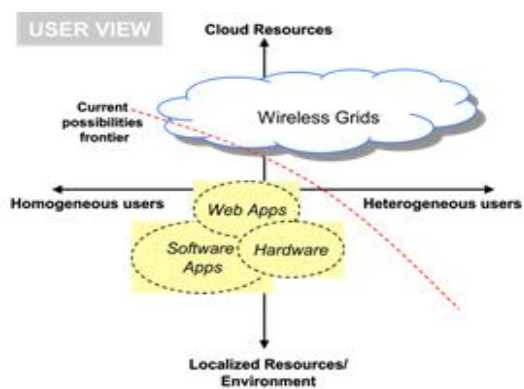
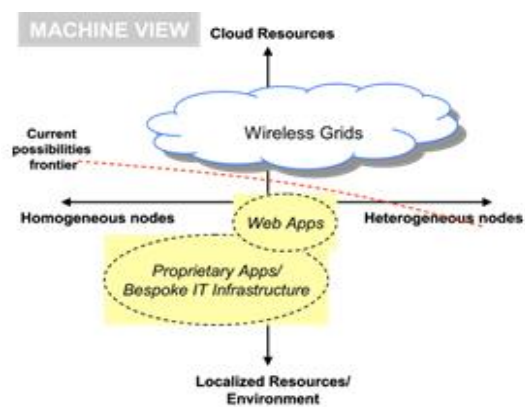

Figure 5. User View             Figure 6. Machine View
(Source: McKnight, ed. WiGiT v0.3)    (Source: McKnight, ed. WiGiT v0.3)

The vision of the IoT will be driven by FMEC, wireless grids, Cloud computing and other various technologies, which also include improving information reliability and efficiency and enhancing customer participation. Given the obvious benefits of the IoT, its introduction has also posed severe security concerns. As a critical infrastructure, the IoT is expected to be a tempting target for hacking, service theft, sabotage, terrorism and other malicious attacks (Li et al. 2012). IoT security has been widely recognized as a major issue with potentially catastrophic implications. Due to its heavy reliance on the cyber-infrastructure for sensing and control, the IoT will be exposed to new risk from computer network vulnerabilities as well as inherit existing risks from physical vulnerabilities within existing systems (Li et al. 2012). Because of the significant role of the IoT as a key infrastructure, the cyber-attacks against it pose sever threats to the security of the architecture.

Both cascading failures and collapses are catastrophic events and will finally lead to large-scale shutdown of wireless grids, FMECs and IoT architectures. Therefore, the authentication of ICD's will be key for wireless grids for the IoT architecture. An authentication model is needed to analyze the wireless grid ICD's for the wg-IoT architecture. In general, the framework provides a generic process for understanding the authentication and authorization component (AAC), the messaging and presence component (MPC) and the security component (SC) from the RSP (McKnight et al. 2015). In this chapter, the consideration is that the proposed framework can be easily extended to analyze a coordinated authentication cyber-attacks launched by attacker's trying to gain access to the overall IoT architecture.

## 6. CONCEPTUALIZATION OF AUTHENTICATION IN A FOG-MOBILE EDGE COMPUTING ENVIRONMENT UTILING THE WIRELESS GRID RESOURCE SHARING PROTOCOL

Authentication is an important issue for the security of fog computing since services are offered to massive-scale end users by front fog nodes (Yi et al. 2015, August). Amor et al. (2017) introduce of a mutual authentication between Fog users at the Edge of the network and the Fog servers at the Fog layer proposing a fog user-fog server anonymous mutual authentication scheme; in which the fog user and fog server authenticate each other and establish a session key without disclosing user's real identity. This scheme is based on Pseudonym Based Cryptography (PBC), Elliptic Curve Discrete Logarithm Problem (ECDLP) and bilinear pairing to establish the session key (Amor et al. 2017). Alharbi et al. (2017) propose a Fog Computing-based Security (FOCUS) system, which leverages a virtual private network (VPN) to secure the access channel to IoT devices through a challenge-response authentication process to protect the VPN server against distributed denial of service (DDoS) attacks. Furthermore, as the emergence of biometric authentication, such as fingerprint authentication, face authentication, touch-based or keystroke-based authentication etc, in mobile computing and cloud computing, applying biometric-based authentication in fog computing will be beneficial (Yi et al. 2015, June).

Authentication requires that an ICD prove its identity. Authentication will require that the ICD authenticate to the IoT network, and the network should authenticate the ICD. The wireless grid core components must be embedded in IoT devices or sensors depending on their capability to handle five primary functions: (1) management of identification [ID] and presence, (2) permissions management, (3) data transfer ability, (4) application programming interface [API] and (4) security (McKnight et al. 2015). These are the elements that make the grid-enabled ecosystem possible and make every ICD a node on the wireless grid. The layers above the core are comprised of the wireless grid's API, which enables connections with other applications and services, the grid user interface (GUI), which may or may not be necessary depending on the thing/object upon which it resides, and finally the wireless grid 'Edgeware' applications, which are typically delivered as services (McKnight et al. 2015) of resource sharing and functionality services and once a wireless grid is established, then resources and services can be published or accessed across the grid, enabling the infinite functional possibilities of wireless grid technology (McKnight et al. 2015).

The wireless grid is made possible by the 'Grid Core', which is software installed on any Grid-enabled, IoT device (McKnight et al. 2015). As identified in Figure 7, the resource sharing protocol (RSP) is the primary Grid Core function provided by the architecture's eight core components: AAC, BAC, MPC, MC, RMC, ELP, CPC and SC (McKnight et al. 2015). The RSP enables creating, joining and subscribing to a wireless grid through provision of the following services: resource identification, resource acquisition, resource advertisement/discovery, communication amongst wireless grids, communication with the internet, creating a wireless grid and joining and subscribing to a wireless grid (McKnight et al. 2015).

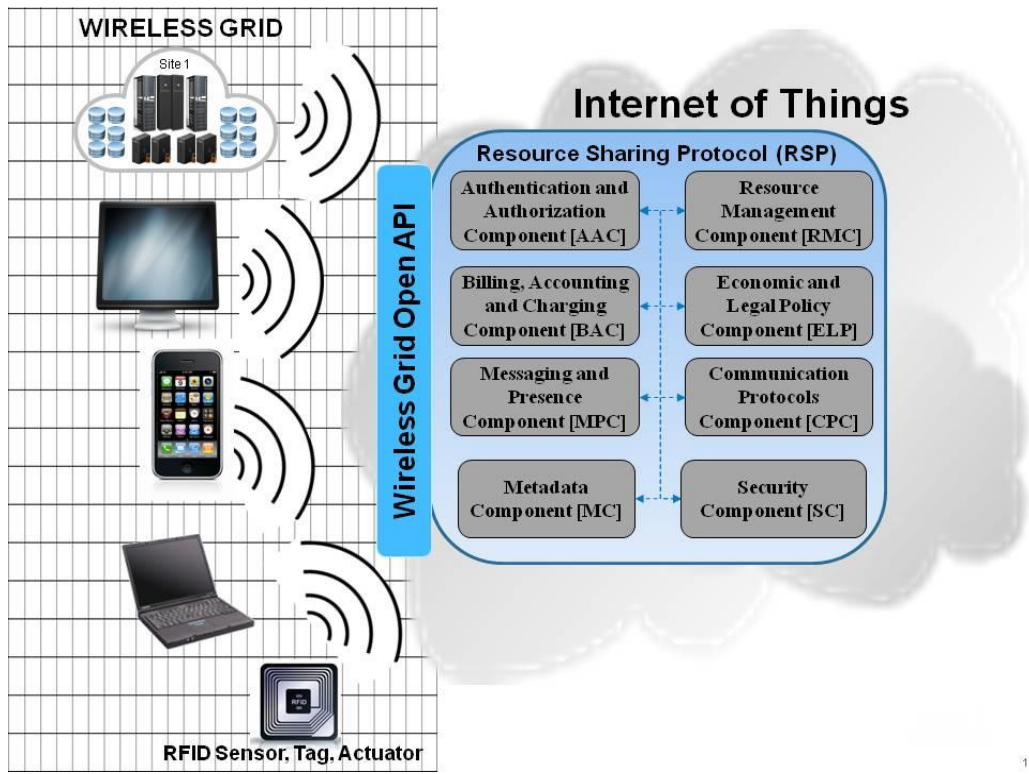

Figure 7. Open Wireless Grid API Map

## 7. WIRELESS GRID INFORMATION ELEMENT (WGIE)

Due to the processing nature of wireless grids for the IoT architecture (see Figure 8), a standard process for authentication needs to be defined that provides for a network-wide cryptographic challenge and response mechanism. This standard should provide for the ability to uniquely authenticate ICD's to legitimate mobile access points. In order to provide authentication for the ICD's in the IoT, the system must use standard cryptographic algorithms/ciphers for wireless and RFID systems (e.g. WPA2, Simon and Speck, AES, Skipjack, etc.) and the establishment of an information element known only to the service provider from the RSP. This information element is referred to as the 'Wireless Grid Information Element [WGIE]', should be programmed into all mobile access points and known to the network infrastructure for the FMEC/IoT network.

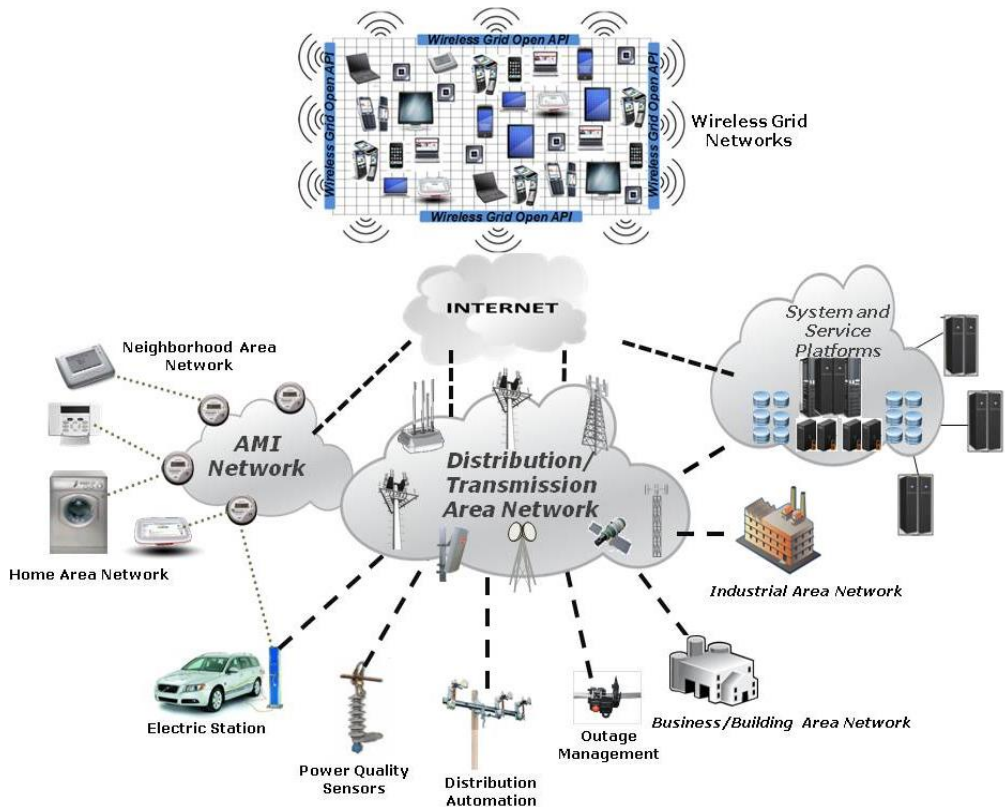

Figure 8. IoT Architecture
(Source: Eichhorn 2010)

The WGIE, as displayed in Figure 9, is a 256-bit cryptographic key variable stored in the semi-permanent memory of the mobile access point and is known to the wireless base register authentication center (WBRAC) of the IoT system.

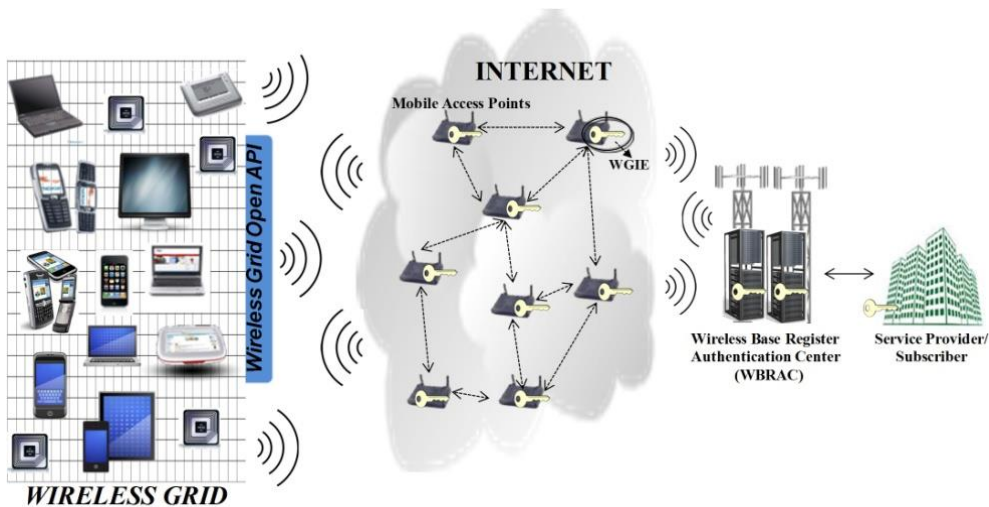

Figure 9. WGIE

The WBRAC provides identification, authentication and encryption for ICD's, provides a central location for supporting mobile access points and is the repository for all subscriber information

As a secured database that stores essential encrypted subscriber information, the WBRAC plays a key role in the operation of the wg-IoT network as it manages the services to which an ICD has subscribed. The WBRAC is constantly subjected to heavy load of traffic to identify service providers/subscribers, route, handle data, and provide security for the IoT system. The WBRAC is used to conduct the IoT network's primary security functions and it contains the wg-IoT network's set of algorithms necessary to authenticate an ICD.

## 8. THE 'WG-IOT' AUTHENTICATION PROCESS

The wg-IoT authentication process (as displayed in Figure 10) is initiated when a mobile access point(s) attempts to confirm the identity of an ICD. This is based on the services identified from the Grid Core and generated in the wg-IoT system. Within the WGIE, the wg-IoT service data (IoT_SD) is a 128-bit pattern that resides in the ICD's semi-permanent memory. The IoT_SD is passed wirelessly across the air between the ICD and the mobile access point. The WGIE includes the 64-bit mobile access point's ESN and the ICD_IN, which is also stored in the WBRAC. The IoT_SD is subdivided into two 64-bit potions called the IoT_SD_1 and the IoT_SD_2. The IoT_SD_1 supports the ICD's authentication procedure and the IoT_SD_2 supports privacy and message confidentiality. When the ICD sends a secure activation signal to the wg-IoT system, the ICD uses the WGIE (i.e. IoT_SD field, ESN, ICD_IN) to begin the generation procedure for the security component (SC) authentication key [SC_auth-K] (e.g. the SC_auth-K is a 128-bit long sequence that is stored in the ICD permanent security identification memory). This information is used to execute the authenticate-signature procedure from the Grid Core's RSP, which will yield a 128-bit numerical value for AAC.

The ACC component handles the authentication of the ICD and the authorization of resources; in effect, the AAC provides the protocols to identify the ICD and understands that ICD's relationship to a resource (i.e. what the ICD can or can't do with a resource). The AAC is used by all grid members and has an identity system that looks at all ICD's and allows policies to be made regarding the ICD's grid profile. The ICD then combines the AAC with two more numerical values from the Grid Core, the messaging and presence component [MPC] and the resource management component [RMC].

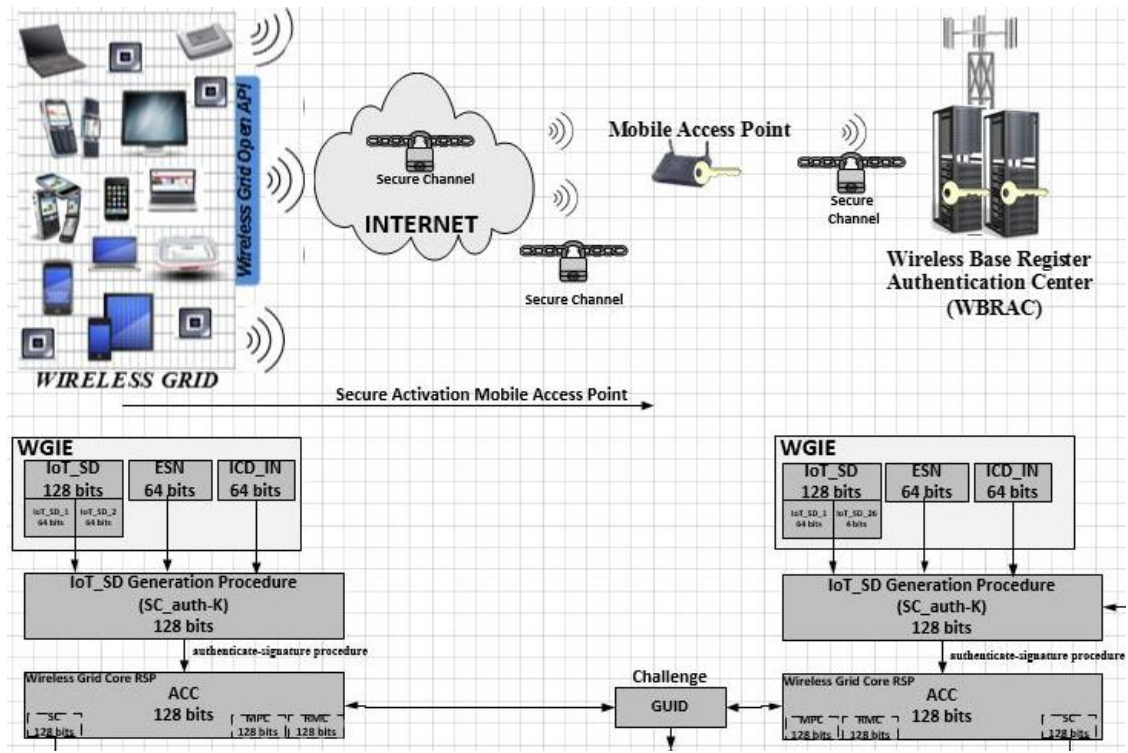

Figure 10. wg-IoT Authentication

The MPC is a scalable messaging and presence protocol that manages the availability of an ICD and the method or language of communication with that resource. The RMC is responsible for aggregating and searching metadata about resources within the context of authentication and works closely with the AAC and MPC. The RMC has a scheduler to manage and coordinate resources, such as network access, and allows for ICD's to be identified as being available. The MPC and the RMC are filled in from values already residing in the ICD's memory. The MPC is a 128-bit long value that was received by the thing/object during its last access parameter message transmitted by the network over the wg-IoT communication channel. The MPC is periodically generated and issued by the wg-IoT system and, it is transmitted by the network to all mobile wireless base stations. The ICD stores and uses the most recent version of the MPC that it has in its authentication attempts. The RMC is a 128-count field stored in the ICD and updated whenever a parameter update order is received from the wg-IoT network on the wg-IoT secure traffic channel. After the ICD has assembled the AAC-MPC-RMC number, called the global unique identifier (GUID), it sends it to the mobile access point. The mobile access point compares the information it received from the ICD with its stored value for the AAC and MPC, along with its RMC derived from a stored service. If valid, the ICD is authenticated. If any of the comparison fails, the mobile access point initiates a unique challenged response process or starts an update value process.

In the update value process (see Figure 11), the wg-IoT system responds by calculating new authentication values and challenging the ICD. First, the WBRAC in the wg-IoT system sends an update message to the mobile access point telling it that the ICD is to be updated and it passes a value to the mobile access point. This random number that the WBRAC has used, together with the ICD to calculate the new value. Since the WGIE is stored in both the ICD and in the WBRAC, the mobile access point sends the value to the ICD in an update order over the wg-IoT secure traffic channel. When the update order arrives at the ICD, it causes the unit to execute an IoT_SD generation procedure. In this procedure, the ICD uses the AAC, ESN and the SC_auth-K to produce an IoT_SD_New value. Meanwhile, the WBRAC has also used the AAC and SC_auth-K to generate its version of the IoT_SD_New value. Next, the ICD generates a random, 256-bit long number called the "thing-object mobile access point (TO_MAP)". The TO_MAP number is used to challenge the mobile access point. The mobile access point issues the challenge via a mobile access challenge order message that is transmitted via the wg-IoT secure traffic channel. The ICD continues on by running an authorization-signature procedure using its IOT_SD_New and the TO_MAP (coupled with the ESN and ICD_IN). The result is a new number called the AUTH_SIGN_MAP within the wireless grid security component.

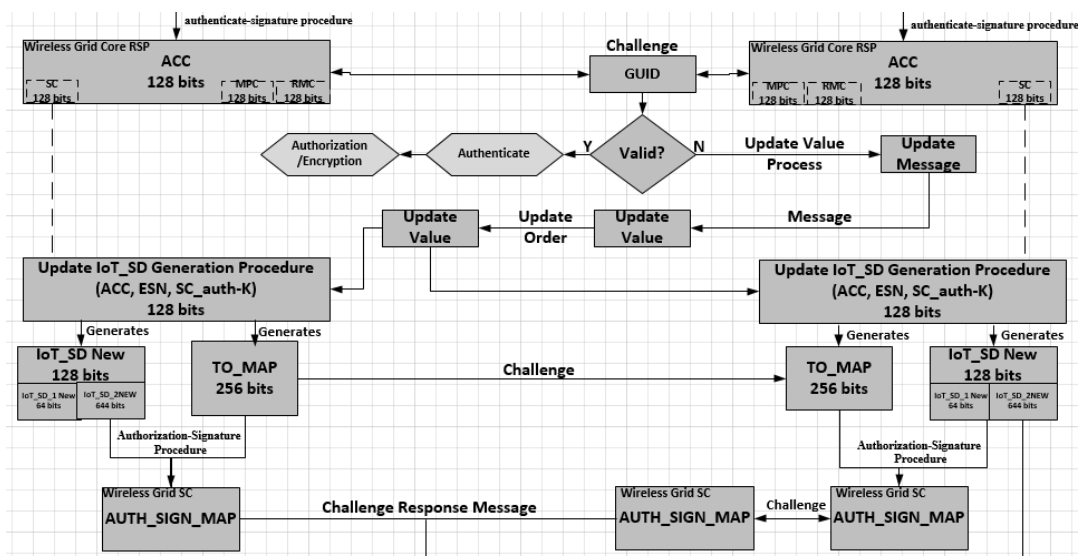

Figure 11. Update Value for AUTH_SIGN_MAP

At the same time that the ICD is calculating the AUTH_SIGN_MAP, the mobile access point received the TO_MAP from the ICD and has started its own authorization-signature procedure. The mobile access point forwards the challenge order to the WBRAC and sends an acknowledgement on receipt of the order to the ICD. The SC uses the TO_MAP (calculated by the ICD) and the IOT_SD_New (calculated by the WBRAC) to generate its own version of the AUTH_SIGN_MAP. The SC then sends its AUTH_SIGN_MAP to the mobile access point in a mobile access point challenge response message. The mobile access point sends this response to the ICD via a mobile access point challenge response order message via the wg-IoT secure traffic channel. The mobile access point challenge confirmation order must be received by the ICD within a timer period of 1s after it has received acknowledgement of reception of the challenge order. If it does not receive the confirmation order within the specific time, the ICD will discard the AUTH_SIGN_MAP and terminate the update value process.

Once the ICD has both copies of the AUTH_SIGN_MAP in its possession, it conducts a comparison of the two. If the comparison results in a successful match, the ICD resets the original IOT_SD_New, replacing them with the new values of the IOT_SD_1New and the IOT_SD_2New. If the comparison results in an unsuccessful match, the ICD discards the IOT_SD_1New and the IOT_SD_2New values and sends an update rejection message to the mobile access point; thus, denying the wg-IoT system access and causing the authentication procedure to begin again. With a successful match and reset, the ICD sends an update confirmation message to the mobile access point. After it has received this message, the mobile access point resets its IOT_SD_1 and the IOT_SD_New value using the IOT_SD_1 New and the IOT_SD_2 New values it received from the WBRAC. The ICD has now been authenticated and proceeds to the authorization and encryption process (see Figure 12).

If any of the calculated values fails a comparison made by the mobile access point during an authentication procedure, then the mobile access point may deem the attempt unsuccessful. It can then terminate the current authentication procedure and initiate the unique challenge response process. This process can be carried out on the wg-IoT secure traffic channels since the mobile access point generates a 64-bit long value called the WMAP and send it to the ICD via the authentication challenge message. This message is sent on the wg-IoT secure traffic channel. When the ICD receives this message, it performs an authorization-signature calculation. Its takes the value and makes it the 64 most significant bits of the WMAP. It also takes the 16 least significant bits of the WBRAC and makes those the 16 least significant bits of the WMAP. The ICD then calculates the authorization-signature, which is used to fill the AUTH_SIGN_MAP field.

The AUTH_SIGN_MAP is then sent to the mobile access point. The mobile access point then compares the ICD AUTH_SIGN_MAP with its own version of the AUTH_SIGN_MAP. If the two do not match, the mobile access point will drop the attempt, deny any further attempt to access the wg-IoT network by the ICD or initiate an update process. If authentication is successful, the mobile access point can move on to authorization and message encryption, which involves scrambling the data signal stream within the wg-IoT network as displayed in Figure 12.

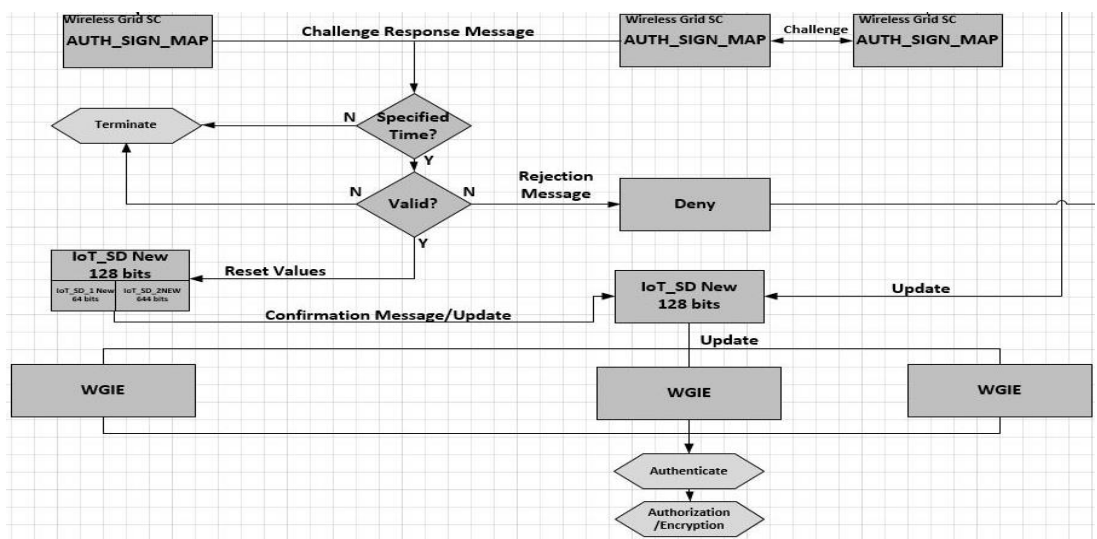

Figure 12. Validating the AUTH_SIGN_MAP

## 9. CONCLUSION

Unlike wired networks, wireless grids and FMECs are not limited by physical space. This potentially opens up the network to attack from rogue ICDs and users who may spy on the wireless transmission or gain unauthorized access to the network from the inside or outside. Traditional thieves, hackers, high-tech criminals, government sponsored organizations, viruses, and other types of malicious code will continue to be causes for security concerns on these networks. The information passed on these networks is exposed to malicious attempts to obtain the information without proper authorization. These new systems must be designed to minimize the vulnerabilities of the networks and the information contained within them. However, as vulnerable as wired networks are for potential attacks, these networks are even more vulnerable.

The intent of this article is to hypothesize a generic authentication methodology for wireless grid/FMECs use for the IoT architecture. The proposed methodology, called wg-IoT, includes the integration of fog computing, wireless grids and FMECs to create this new IoT architecture and must be further researched. The authentication process developed from the wireless grid RSP still has to be developed and proposed for modeling the authentication of ICD's, which allows strategies and approaches for enhancing the information security architecture for further development. There are numerous complex considerations which must be considered when implementing this process and without adequate forethought these new ICDs and FMECs may be ill-advised.

International Journal of Ubiquitous Computing (IJU), Vol.13, No.1/2, April 2022